\documentclass[10pt,conference]{IEEEtran}
\pdfoutput=1
\usepackage{graphicx}
\usepackage{amsmath}
\usepackage{tabularx}  
\usepackage{multirow}  
\usepackage{booktabs}  
\usepackage{caption} 
\usepackage{url}
\usepackage{algorithm,algpseudocode}
\usepackage{amsfonts}  
\usepackage{hyperref}

\author{
\IEEEauthorblockN{Pol G. Recasens\IEEEauthorrefmark{1}\IEEEauthorrefmark{2}, Ferran Agullo\IEEEauthorrefmark{1}\IEEEauthorrefmark{2}, Yue Zhu\IEEEauthorrefmark{4}, Chen Wang\IEEEauthorrefmark{4}, \\ Eun Kyung Lee\IEEEauthorrefmark{4},  Olivier Tardieu\IEEEauthorrefmark{4}, Jordi Torres\IEEEauthorrefmark{2}\IEEEauthorrefmark{3}, Josep Ll. Berral\IEEEauthorrefmark{3}\IEEEauthorrefmark{2}}
\IEEEauthorblockA{\IEEEauthorrefmark{2}Barcelona Supercomputing Center, \{pol.garcia, ferran.agullo, jordi.torres\}@bsc.es}
\IEEEauthorblockA{\IEEEauthorrefmark{3}Universitat Politècnica de Catalunya - BarcelonaTech (UPC), \{josep.ll.berral\}@upc.edu}
\IEEEauthorblockA{\IEEEauthorrefmark{4}IBM Research, \{Yue.Zhu, Chen.Wang1, eunkyung.lee, tardieu\}@us.ibm.com}}

\title{Mind the Memory Gap: Unveiling GPU Bottlenecks in Large-Batch LLM Inference}

\begin{document}
\maketitle

\renewcommand{\thefootnote}{*}
\footnotetext{denotes equal contribution.}
\renewcommand{\thefootnote}{\dag}

\begin{abstract}
Large language models have been widely adopted across different tasks, but their auto-regressive generation nature often leads to inefficient resource utilization during inference. While batching is commonly used to increase throughput, performance gains plateau beyond a certain batch size, especially with smaller models, a phenomenon that existing literature typically explains as a shift to the compute-bound regime. In this paper, through an in-depth GPU-level analysis, we reveal that large-batch inference remains memory-bound, with most GPU compute capabilities underutilized due to DRAM bandwidth saturation as the primary bottleneck. To address this, we propose a Batching Configuration Advisor (BCA) that optimizes memory allocation, reducing GPU memory requirements with minimal impact on throughput. The freed memory and underutilized GPU compute capabilities can then be leveraged by concurrent workloads. Specifically, we use model replication to improve serving throughput and GPU utilization. Our findings challenge conventional assumptions about LLM inference, offering new insights and practical strategies for improving resource utilization, particularly for smaller language models. The code is publicly available
at \url{https://github.com/FerranAgulloLopez/vLLMBatchingMemoryGap}.
\end{abstract}

\section{Introduction}

Large language models have been traditionally designed as a single general-purpose architecture. However, the rise of agentic AI has shifted interest towards smaller, specialized LLMs designed for domain-specific tasks and collaborative, distributed executions. These models offer greater flexibility, adapting efficiently to various applications while requiring significantly fewer computational resources. Smaller models can achieve competitive performance in domains such as mathematics and code generation~\cite{liu2024deepseek, team2024gemini} through post-training optimizations like knowledge distillation from general-purpose models~\cite{guo2025deepseek}. Also, recent advances in serving optimizations have made these models accessible to resource-limited users. However, their smaller size introduces unique inference challenges. Unlike larger models, which are constrained by high memory demands, smaller models face different performance plateaus and require careful configuration to optimize performance without over allocating non-bottleneck resources.

Serving a user request with a LLM can be divided into two distinct phases: the \textit{prefill} phase and the \textit{decode} phase. During prefill, the model processes all input tokens in parallel and generates the first output token, efficiently using computational resources. In contrast, the decode phase generates one token at a time in an auto-regressive manner, leading to large memory transfers despite low computational demand. Batching requests~\cite{kwon2023efficient, yu2022orca} improves resource utilization in this phase by generating multiple output tokens per forward pass, enhancing serving throughput. To minimize waiting times for batch completion, most recent schedulers dynamically determine which requests join or leave the batch per forward pass \cite{yu2022orca}, avoiding unnecessary stalls.

Due to GPU memory constraints, batching a large number of requests is often impractical. However, for smaller models, hundreds or even thousands of requests can fit in a single GPU, improving throughput up to a knee-point, beyond which batching additional requests yields diminishing throughput returns and increases latency~\cite{kwon2023efficient}. Arithmetic intensity\textemdash the ratio between compute operations and memory bytes accessed\textemdash indicates whether a workload is memory-bound or compute-bound. No-batch inference has been well established as memory-bound, where the memory transfer time of model weights and the KV cache exceeds computation time~\cite{agrawal2024taming}. While some studies~\cite{yuan2024llm, agrawal2024taming} assume that large-batch LLM inference transitions to the compute-bound regime, this assumption has never been rigorously validated through an in-depth GPU-level analysis. As a result, the misconception persists that large batches fully utilize GPU resources~\cite{agrawal2024taming, yuan2024llm}.

\begin{figure}[ht]
  \centering
  \vspace{-5pt}
  \includegraphics[width=0.68\linewidth]{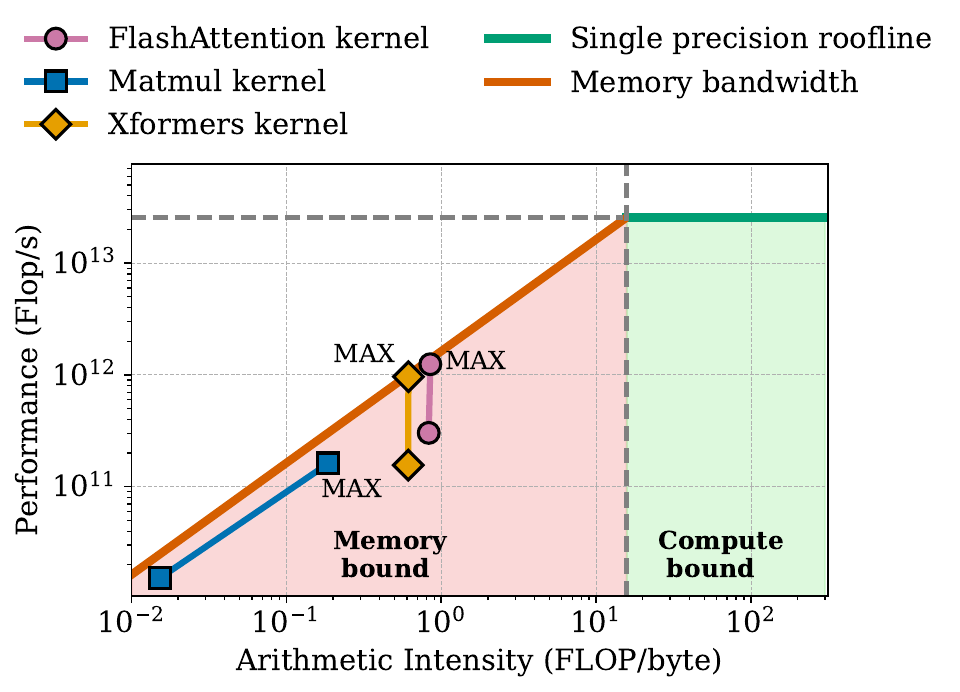}
  \caption{Performance vs Arithmetic intensity of attention and matrix multiplication kernels for batch size 1 and the maximum batch size (MAX). While batching increases the arithmetic intensity of matrix multiplications, the arithmetic intensity of xFormers and FlashAttention attention kernels—two memory-optimized attention implementations—remains nearly constant, leading to DRAM saturation. The data was extracted using NVIDIA Nsight Compute from the last decode step of OPT-1.3B on an H100 GPU.}
  \label{fig:roofline}
\end{figure}

In this work, we conduct a detailed GPU analysis to uncover the true causes of the throughput plateau in large-batch LLM inference. Our findings reveal that the primary performance bottleneck during decoding stems from the attention mechanism. Specifically, we identify DRAM bandwidth saturation as the main limiting factor, with over 50\% of the attention kernel cycles stalled due to data access delays for all tested models. Figure~\ref{fig:roofline} presents the arithmetic intensity for two attention implementations as batch size increases from 1 to the maximum allowed by GPU memory (MAX). Our results clearly demonstrate that the key components of the decoding step\textemdash attention and matrix multiplication kernels\textemdash remain deep within the memory-bound regime across all batch sizes. While the matrix multiplication (matmul) kernels gain arithmetic intensity as the batch size grows, the arithmetic intensity of both attention kernels remains nearly constant. 

Due to GPU DRAM bandwidth saturation, large batch sizes consume substantial GPU memory without yielding proportional throughput gains and significantly degrading latency. To mitigate these issues, we also propose a Batching Configuration Advisor (BCA), a profiling-driven method to determine the optimal batch size $B_{opt}$ considering both the throughput plateau and a user-defined latency. By identifying a balanced batch size, BCA minimizes GPU memory waste which can be leveraged by concurrent workloads, similar to CPU resource management in cloud environments. Model replication\textemdash running multiple LLM instances concurrently on the same GPU\textemdash overlaps GPU operations and increases resource utilization, mitigating the found bottlenecks. Specifically, compared to using a single model replica with MAX memory allocation, model replication increases throughput by 33.7\% for OPT-1.3B, and 7.49\% for OPT-2.7B. 


In summary, our key contributions are:

\begin{itemize}
    \item \textbf{We demonstrate that LLM inference remains memory-bound even at large batch sizes}.  DRAM bandwidth is the limiting factor in large-batch regimes, with over half of attention computation cycles stalled due to memory access delays.
    \item \textbf{We propose a Batching Configuration Advisor (BCA)}, which recommends an optimal batch size and memory allocation to avoid the throughput plateau while adhering to latency constraints. 
    \item \textbf{We evaluate the benefits of freeing GPU memory with BCA} by reallocating freed resources to serve concurrent model replicas. This approach increases GPU resource utilization and substantially improves overall throughput by overlapping operations and mitigating idle times.
\end{itemize}

\section{Background}

\subsection{Autoregressive Generation}

Decoder-only language models such as OPT~\cite{zhang2022OPT}, GPT~\cite{brown2020language}, and Llama~\cite{touvron2023Llama}, excel in language comprehension and generation, often demonstrating strong zero-shot capabilities across various tasks. These models are typically trained on next-token prediction, where tokens are generated autoregressively based on a given input prompt $x$. Model parameters are typically optimized by minimizing the negative log-likelihood loss of the predicted token probabilities: 
\begin{equation}
    P(x_{n+1}|x_1,...,x_n)
\end{equation}

Built upon the transformer architecture~\cite{vaswani2017attention}, these models consist of stacked blocks with self-attention and feed-forward layers. In each block, the self-attention module identifies relevant tokens in the input sequence, modeling relationships between tokens and capturing both long and short-range dependencies. To achieve this, the input tensor is linearly transformed using learnable matrices into keys $K$, queries $Q$, and values $V$ matrices, which compute attention scores to quantify token importance in a highly parallelizable manner. Standard attention kernels perform HBM accesses quadratic in sequence length \cite{dao2022flashattention}, and most of its operations are memory-bound, such as softmax and dropout, which require frequent memory transfers with minimal computation per element. Kernel fusion techniques mitigate this limitation by combining multiple operations within the same kernel, reducing redundant memory accesses. For instance, FlashAttention \cite{dao2022flashattention} employs tiling to fuse all attention operations in one CUDA kernel, significantly reducing memory accesses.

In LLM inference, processing a new input sequence begins with the \textit{prefill} phase, where all input tokens are processed in parallel to generate an initial output token. Next, the model transitions into the \textit{decode phase}, during which each subsequent token is generated autoregressively\textemdash one at a time\textemdash by conditioning on previously generated tokens. This stage ends when the model generates an end-of-sequence token or reaches the maximum output length. To avoid redundant recomputations, intermediate results\textemdash specifically the key and value pairs\textemdash are stored in GPU memory as the KV cache. This cache enables efficient reuse of computed attention states from previous tokens and reduces the attention score computation from a matrix-matrix product $QK^T$ to a matrix-vector product $qK^T$, where $q$ is the query vector for the current token. In contrast to the prefill phase, the decode phase involves significant memory transfers of key-value pairs and model weights relative to the minimal computations performed. This disparity creates a primary bottleneck in LLM inference, driving the need for optimizations to improve both latency and throughput.

\subsection{Memory vs. Compute Performance Limitations}

The performance of a compute operation can be decomposed into two primary components: \textbf{memory time $T_{M}$}, the time spent transferring data from HBM to the on-chip SRAM (including model weights and KV values), and \textbf{compute time $T_{C}$}, the time spent computing the arithmetic operations. Whether the operation is memory-bound or compute-bound is determined by its \textbf{arithmetic intensity}, defined as the ratio of FLOPs to bytes accessed from memory. A low ratio indicates a memory-bound regime where memory accesses dominate, while a high ratio signifies a compute-bound regime, where compute operations govern. Ideally, optimal resource utilization occurs when $T_{C} = T_{M}$.

In LLM inference, this classification can be applied to different inference stages. The prefill phase is known as compute-bound, due to parallelized computations, while the decode phase becomes memory-bound due to frequent memory accesses, the sequential nature of token generation, and limited parallelism \cite{agrawal2024taming}. This coarse-grained perspective holds for no-batch inference and explains the resource efficiency plateaus observed during decoding. However, this simplified view often leads to the assumption that the observed performance plateau at larger batch sizes indicates a transition to the compute-bound regime~\cite{agrawal2024taming, yuan2024llm}.

\subsection{Throughput-Latency Trade-Off}

Previous scheduling systems~\cite{agrawal2024taming, kwon2023efficient} introduced various policies for managing the prefill and decode phases of incoming requests, aiming to maximize throughput while maintaining low latency. Recent optimizations, such as chunked prefill~\cite{agrawal2024taming}, further enhance efficiency by combining both phases into the same forward pass. These schedulers primarily rely on batching to address the low compute utilization during decode. While achieving sufficiently large batch sizes for LLMs is typically impractical due to excessive memory demands, the shift toward smaller, specialized models\textemdash combined with optimizations that reduce memory and computational requirements\textemdash makes it both feasible and important to study the impact of larger batch sizes on smaller models in resource-constrained environments. This impact, however, is not always positive, as after a certain batch size throughput improvements begin to plateau, resulting in diminishing returns \cite{recasens2024towards}.

\begin{figure}[ht]
\centering
  \vspace{-5pt}
  \includegraphics[width=0.8\linewidth]{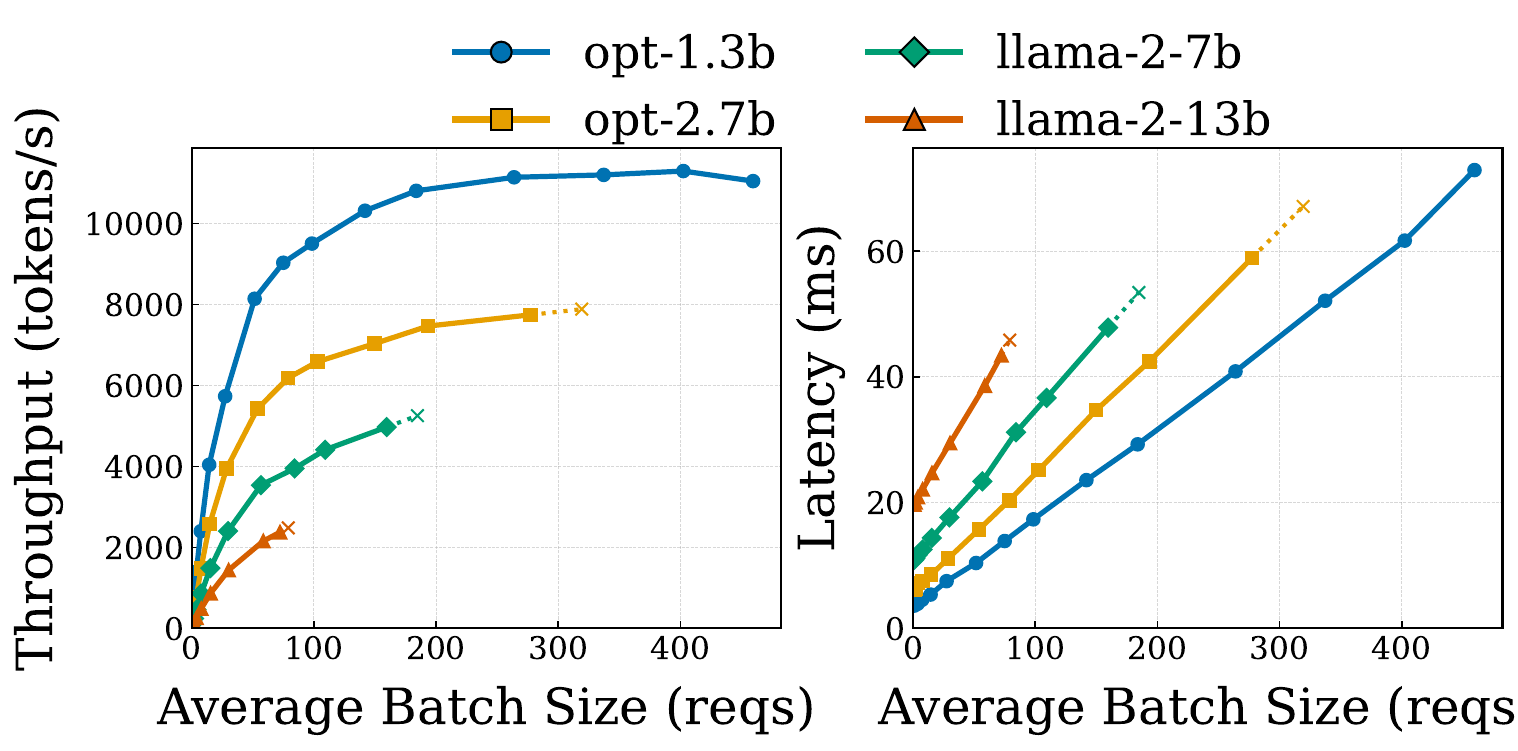}
  \caption{Throughput (input and output tokens/s) and latency (inter-token latency) evolution when setting the maximum batch size to values in range 1..512 across different models (OPT-1.3B, OPT-2.7B, Llama-2-7B and Llama-2-13B). The X-axis corresponds to the average batch size, instead of the set maximum, and the crosses mark the point where the KV cache capacity is exceeded due to the increased batch size. Results are obtained through the online mode described in Section~\ref{sec:methodology}.}
  \label{fig:background_throughput_latency}
\end{figure}

\begin{figure}[ht]
  \centering
  \includegraphics[width=0.7\linewidth]{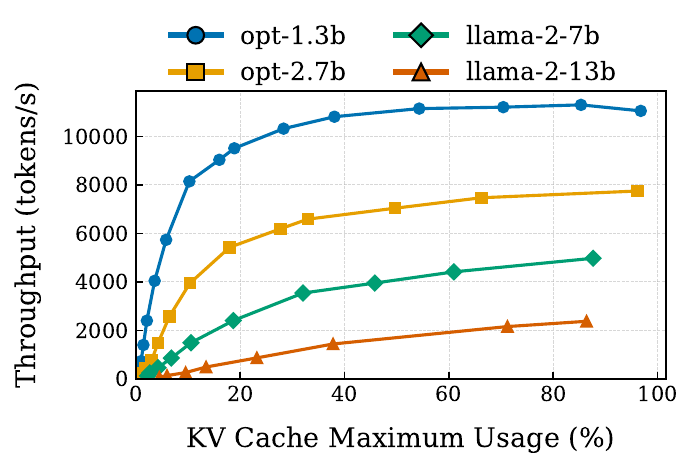}
  \caption{Comparison between throughput (input and output tokens/s) and the maximum KV cache usage when setting the maximum batch size to values 1..512 across different models (\textit{OPT-1.3B}, \textit{OPT-2.7B}, \textit{Llama-2-7B} and \textit{Llama-2-13B}). Results are obtained through the online mode described in Section~\ref{sec:methodology}.}
  \label{fig:background_kv_cache}
\end{figure}

To contextualize this throughput-latency trade-off, Figure~\ref{fig:background_throughput_latency} illustrates the observed throughput plateau, demonstrating how increasing batch sizes beyond a certain knee-point results in diminishing throughput gains alongside significantly increasing inter-token latency. On the other hand, Figure~\ref{fig:background_kv_cache} shows that the memory required to reach this performance plateau is only a fraction of the KV cache. For instance, OPT-1.3B achieves almost maximum throughput using just 40\% of its KV cache, while OPT-2.7B requires 50\%. Increasing batch size further yields only marginal throughput gains, at the cost of a larger GPU memory usage. Since current serving frameworks automatically allocate the maximum possible memory without accounting for performance plateaus or diminishing returns, this potentially leads to inefficient resource utilization.

\section{Related Work}\label{sec:relatedwork}

\subsection{Serving Language Models} 

As language models continue to advance the state-of-the-art across diverse tasks, efficiently deploying and serving these models has become a critical area of research. This has led to the development of several serving systems such as Orca~\cite{yu2022orca}, Text Generation Inference (TGI)~\cite{hf-tgi}, DeepSpeed-FastGen~\cite{aminabadi2022deepspeed,deepspeed-fastgen}, Sarathi-Serve~\cite{agrawal2024taming}, and vLLM~\cite{kwon2023efficient}. Complementary approaches, such as AlpaServe~\cite{li2023alpaserve}, focus on multi-model serving by leveraging statistical multiplexing across distributed devices to maximize resource utilization. Additionally, high-performance inference engines like FasterTransformer~\cite{nvidia-fastertransformer} offer C++/CUDA-based implementations with low-level optimizations for further efficiency gains. In this work, we adopt vLLM as our inference framework, as it integrates most recent serving optimizations.

\subsection{Serving Optimizations}
A range of orthogonal optimizations has been proposed to reduce memory usage and improve compute efficiency in LLM inference. \textit{Model quantization}~\cite{xiao2023smoothquant,frantar2022gptq, sheng2023flexgen} lowers parameter precision to reduce memory footprint, often at the cost of accuracy. \textit{Mixture-of-Experts} (MoE)~\cite{shazeer2017outrageously} selectively activates a subset of model parameters during inference, reducing computational cost per decoding step. \textit{Sparsity} techniques~\cite{ma2023llm} prune attention heads or layers to improve efficiency while maintaining performance. \textit{Offloading} techniques such as~\cite{sheng2023flexgen} alleviate on-device memory limitations by leveraging CPU and disk storage for intermediate computations in offline serving scenarios. \textit{Speculative decoding}~\cite{leviathan2023fast} accelerates generation by predicting multiple tokens at once and verifying them in fewer model steps, significantly improving latency. \textit{Multi-query attention} (MQA)~\cite{shazeer2019fast} reduces memory bandwidth and computational overhead by sharing a single key and value head for all query heads. \textit{Grouped-query attention} (GQA)~\cite{ainslie2023gqa} generalizes MQA by dividing the number of query heads in groups, each sharing a single key and value head, achieving accuracy comparable to the standard multi-head attention (MHA). These techniques, often combined, enable faster and more efficient LLM serving while maintaining output quality.

\subsection{Managing the KV Cache}

Managing KV cache in LLM serving is challenging due to the unpredictable number of output tokens per request. This uncertainty makes it difficult to pre-allocate memory efficiently. Initially, serving frameworks such as Orca~\cite{yu2022orca} pre-allocated contiguous GPU memory based on the maximum possible output length. However, this often leads to memory fragmentation for shorter outputs. Recent works have addressed these limitations. $S^3$~\cite{jin2023s} estimates request output length using an auxiliary model, dynamically adjusting memory pre-allocation. \textsc{vLLM}~\cite{kwon2023efficient} introduces PagedAttention, a memory management mechanism inspired by OS paging, which reduces fragmentation by allowing non-contiguous memory allocation. Other approaches focus on reusing precomputed attention states. Prompt cache mechanism~\cite{gim2023prompt} store attention states of frequently visited text segments for reuse across sequences, while RadixAttention~\cite{zheng2024sglang} maintains a radix tree on the CPU to enable KV cache reuse at runtime.

\subsection{LLM Inference Profiling} 

Yuan et al.~\cite{yuan2024llm} identified the decode phase as memory-bound, attributing the bottleneck to frequent memory accesses and characterizing it through the Roofline model, which visually compares operation performance against hardware limits. While they provide a comprehensive evaluation of various inference optimizations and highlight the challenges posed by the memory hardware limit, they assume that large batch sizes result in full compute utilization, without rigorously verifying this through detailed GPU profiling. Similarly, Li et al.~\cite{li2024llm} survey recent advancements in LLM serving that maintain the standard decoding process, but does not take into account the throughput plateau in larger-batch serving. Recasens et al.~\cite{recasens2024towards} empirically showed that large batches lead to a throughput plateau, and hinted potential benefits of model replication. However, this study lacks a detailed GPU tracing to explain the underlying causes of this plateau, leaving the bottlenecks behind throughput saturation in large-batch scenarios unexplored. In this work, we conduct an in-depth GPU analysis that unveils the bottlenecks behind the throughput plateau in large-batch scenarios.  

\section{Methodology} \label{sec:methodology}
We conduct our experiments using the well-established vLLM framework~\cite{kwon2023efficient}, specifically its main branch state as of October 18th, 2024. The framework is configured with default parameters, except for disabling logging, setting the maximum batch size to 4096 tokens, and limiting the maximum context length to 2048 tokens. We employ this framework in two modes: \textbf{online mode} following a client-server architecture, transmitting requests via API endpoints, and an \textbf{offline mode} where all prefill and decode steps are executed directly via Python calls. The online mode is used in Section~\ref{sec:advisor} to evaluate our proposal in a real-world scenario, whereas the offline mode is employed in Section~\ref{sec:gpuprofiling}, allowing a more precise control and analysis of execution phases without additional noise.

\textbf{Hardware.} All experiments are conducted on a single-node setup with an NVIDIA Hopper H100 (64GB HBM2), 128GB RAM memory, and 20 CPU cores.

\textbf{Models}. We evaluate four models: OPT-1.3B, OPT-2.7B, Llama-2-7B, and Llama-2-13B. All models fit within the 64GB GPU,  allowing sufficient memory for large batch processing.

\textbf{Workload}. In online mode, 2000 requests are sampled from a cleaned ShareGPT dataset, maintaining the original input and output length distribution. In offline mode, we generate synthetic requests with fixed input and output lengths. Each request consists of 161 input tokens and 338 output tokens, matching the mean input/output lengths in the original ShareGPT dataset~\cite{sharegpt}.

\section{GPU Profiling and Performance Bottlenecks} \label{sec:gpuprofiling}

In this section, we characterize the \textit{throughput plateau} observed in large-batch regimes and investigate its underlying causes. Unlike prior studies, our work is the first to provide a comprehensive explanation of this performance bottleneck using detailed GPU tracing data. Our analysis primarily relies on two tools: NVIDIA Nsight Systems (2023.2.3) and NVIDIA Nsight Compute (2023.3.0.0). Nsight provides a high-level view of GPU activity across the entire program execution (employed in Sections~\ref{sec:decode_prefill} and~\ref{sec:decode_kernels}), while Nsight offers fine-grained insights into the execution of specific kernels (employed in Section~\ref{sec:attention_kernel}). 

Our findings reveal that DRAM saturation in the attention mechanism is the primary cause of the throughput plateau in large-batch scenarios, challenging prior assumptions that attribute it to a shift toward a compute-bound regime. While batching increases the arithmetic intensity of matrix multiplication kernels, we observe that the arithmetic intensity of attention kernels remains nearly constant. This ultimately leads to memory-bandwidth saturation, leaving a significant portion of computational resources underutilized.

\subsection{Decode vs Prefill} \label{sec:decode_prefill}

\begin{figure}[ht]
  \centering
  \includegraphics[width=0.85\linewidth]{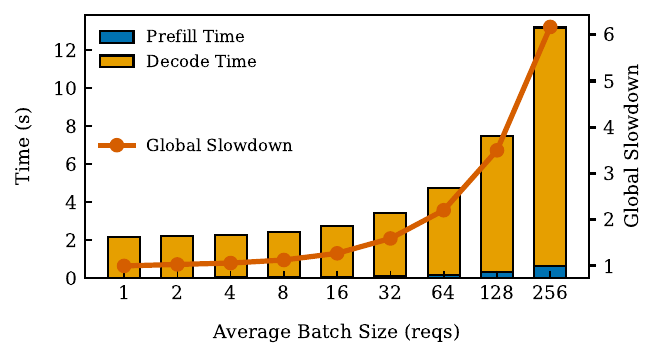}
  \caption{Evolution of total execution time as batch size increases for the \textit{OPT-2.7B} model. We distinguish prefill and decode phases, and the overall slowdown measures the global execution time difference from when running with batch size 1.}
  \label{fig:prefill_decode}
\end{figure}

We begin our analysis with a broad examination of the factors contributing to the throughput plateau, focusing on the evolution of global execution time as batch size increases. Specifically, we distinguish between the prefill and decode phases. In line with prior literature~\cite{agrawal2024taming}, our results confirm that the decode phase is the primary bottleneck in inference serving. As shown in Figure~\ref{fig:prefill_decode}, the decode phase accounts for the vast majority of the total inference time for the OPT-2.7B model in all tested batch sizes. Although prefill becomes relatively more significant at large batch sizes, it remains below 5\% even at the maximum batch size. Table~\ref{table:evaluation_analysis} further corroborates these findings, showing that the decode phase dominates inference time across all tested models under maximum batch conditions.

As shown in Figure~\ref{fig:prefill_decode}, global execution time remains nearly constant until the batch size exceeds 32 requests. Beyond this point, it increases proportionally, leading to a 6x slowdown at the largest batch size. This behavior aligns with the throughput plateau observed in Figure~\ref{fig:background_throughput_latency}, where the throughput of the OPT-2.7B model increases from 225 tokens per second at batch size 1, to 7,607 tokens per second at batch size 256\textemdash an approximate 33.8x increase instead of the expected 256x\textemdash indicating a slowdown of about 7.8x. These results confirm that our offline findings are consistent with online results presented in the background. The slight variation between the two may stem from differences in request length distributions across the two distinct modes.

Table~\ref{table:evaluation_analysis} presents key GPU metrics, categorized into compute-related and memory-related. Modern GPUs consist of multiple Streaming Multiprocessors (SMs), each executing parallel thread groups known as Warps. The results indicate that while most SMs remain active throughout execution\textemdash occasionally reaching full saturation\textemdash their average utilization remains low. Specifically, the \textit{Compute Warps in Flight} metric reveals that no model exceeds 35\% average Warp usage in either the prefill or decode phases. Notably, this metric is higher during prefill, supporting prior claims that this phase is more compute-intensive. The \textit{Unallocated Warps in Active SMs} metric measures the percentage of Warps assigned to an SM but not yet allocated. This value remains consistently high across all models, indicating the presence of a bottleneck preventing Warps from being allocated. Regarding GPU memory read and write operations (denoted as DRAM), write activity is minimal, whereas read values remain consistently high across all models, particularly for larger ones. These high DRAM read values significantly exceed compute percentages, possibly suggesting a memory-bound regime. However, overall GPU resource utilization remains far from saturation at large batch sizes, especially in terms of compute capabilities.

\begin{table*}[!htb]
    \renewcommand{\arraystretch}{1.2}
    \centering
    \begin{tabular}{l|l|l|cc|cc|cc|cc}
    \multicolumn{3}{c|}{} & \multicolumn{2}{c|}{\textbf{OPT-1.3B}} & \multicolumn{2}{c|}{\textbf{OPT-2.7B}} & \multicolumn{2}{c|}{\textbf{Llama-2-7B}} & \multicolumn{2}{c}{\textbf{Llama-2-13B}} \\ \cline{4-11}
    \multicolumn{3}{c|}{} & Prefill & Decode & Prefill & Decode & Prefill & Decode & Prefill & Decode \\ \bottomrule
    \multicolumn{3}{r|}{Importance (\%)} & 0.03 & 0.97 & 0.05 & 0.95 & 0.05 & 0.95 & 0.05 & 0.95 \\ \midrule
    \multirow{6}{5.0em}{GPU compute-related} & \multirow{2}{10.0em}{Active SMs (\%)} & Average & 75.12 & 61.90 & 80.86 & 72.04 & 87.26 & 69.23 & 87.51 & 76.65 \\
    & & Max & 100.00 & 100.00 & 100.00 & 100.00 & 100.00 & 100.00 & 100.00 & 100.00 \\ \cline{2-11}
    & \multirow{2}{10.0em}{Compute Warps in flight (\%)} & Average & 25.88 & 12.91 & 30.08 & 31.14 & 26.84 & 9.85 & 24.61 & 10.27 \\
    & & Max & 96.00 & 100.00 & 96.00 & 97.00 & 91.00 & 90.00 & 92.00 & 72.00 \\ \cline{2-11}
    & \multirow{2}{10.0em}{Unallocated Warps in active SMs (\%)} & Average & 49.24 & 49.00 & 50.77 & 40.90 & 60.42 & 59.39 & 62.89 & 66.39 \\
    & & Max & 88.00 & 88.00 & 82.00 & 82.00 & 88.00 & 88.00 & 88.00 & 88.00 \\ \midrule
    \multirow{4}{5.0em}{GPU memory-related} & \multirow{2}{10.0em}{DRAM read (\%)} & Average & 32.85 & 47.98 & 43.21 & 60.81 & 62.57 & 70.55 & 66.08 & 76.75 \\
    & & Max & 91.00 & 93.00 & 97.00 & 99.00 & 95.00 & 97.00 & 95.00 & 97.00 \\ \cline{2-11}
    & \multirow{2}{10.0em}{DRAM Write Throughput (\%)} & Average & 18.35 & 5.56 & 15.55 & 5.83 & 12.79 & 2.59 & 10.16 & 1.82 \\
    & & Max & 100.00 & 100.00 & 66.00 & 78.00 & 43.00 & 48.00 & 44.00 & 33.00 \\
\end{tabular}
    \caption{Comparison between prefill and decode phases in their relative importance and their results for a selection of key GPU metrics. We set the batch size to the maximum value that fits in KV cache in all tested models, and included the average and maximum value for the full length of the execution in all GPU metrics.}
    \label{table:evaluation_analysis}
\end{table*}

\subsection{Decode Kernels} \label{sec:decode_kernels}

\begin{figure}[ht!]
  \centering
  \includegraphics[width=0.95\linewidth]{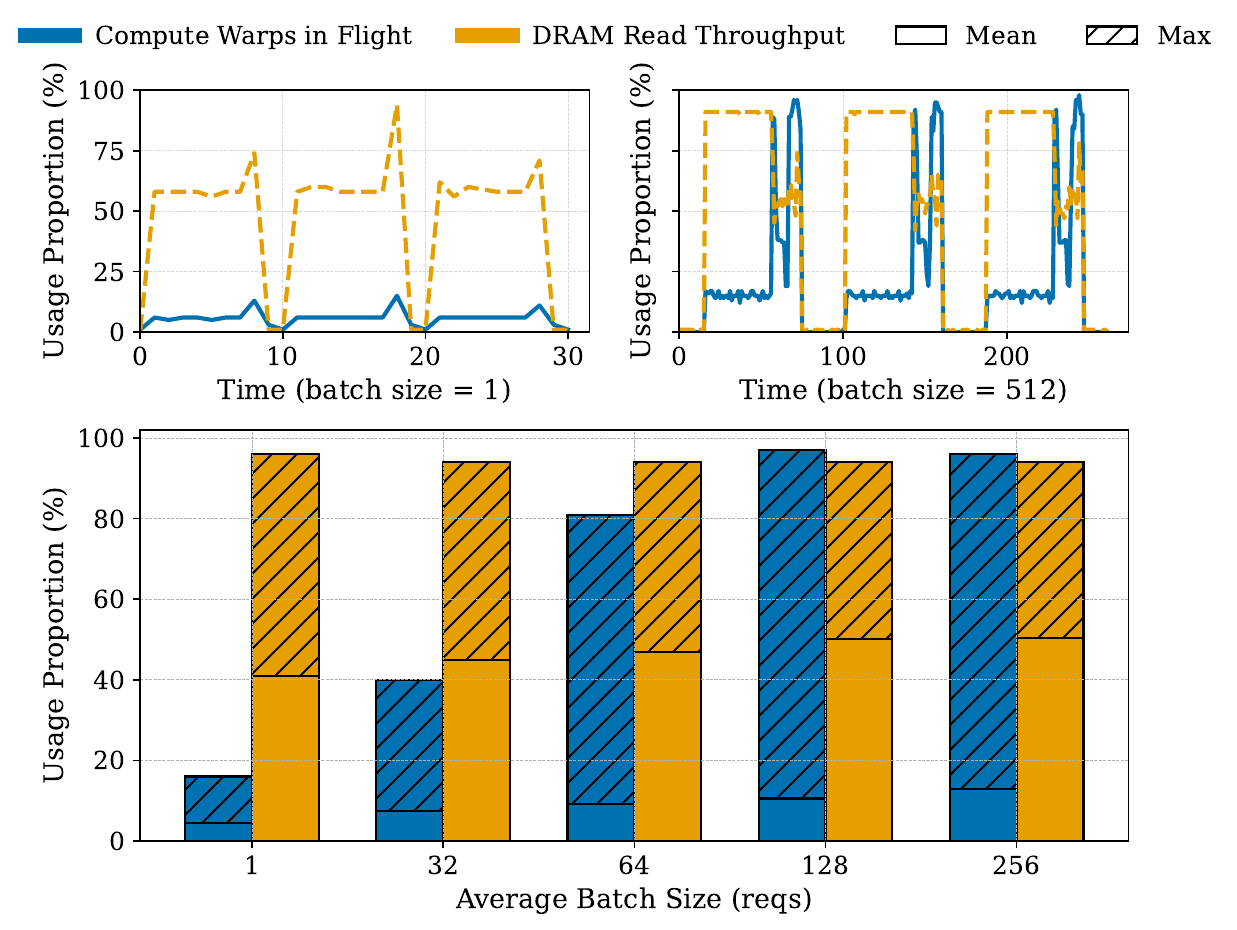}
  \caption{(Top) Evolution of the metrics \textit{Compute Warps in Flight} and \textit{DRAM Read Throughput} in the first three decoding steps of the execution when using model OPT-1.3B with two different batch sizes (1 and 512) (Bottom) Average and maximum values during the full length of the execution of the previous two metrics in five different batch sizes (1, 32, 64, 128, 256 and 512) in OPT-1.3B.}
  \label{fig:decode_kernels_batch_size_evolution}
\end{figure}

In this subsection, we continue our analysis focusing exclusively on the decode phase, as it dominates the execution time. Figure~\ref{fig:decode_kernels_batch_size_evolution} (top) illustrates the evolution of the first three decoding steps when running OPT-1.3B at batch sizes 1 and 512. As shown, DRAM read activity remains consistently high throughout most of each decoding step, while compute utilization stays below 20\%. It is only towards the end that compute usage increases and DRAM read declines. This pattern suggests a memory saturation scenario where Warps stall while waiting for data transfers. Also, there is a noticeable GPU idle gap between decoding steps, which we attribute to CPU processing time. This gap increases with batch size, further impacting overall efficiency. Figure \ref{fig:decode_kernels_batch_size_evolution} (bottom) provides additional insights into GPU resource utilization across multiple batch sizes. While peak DRAM and compute utilization approach saturation, average utilization remains significantly below 50\% at large batch sizes, suggesting the presence of an underlying bottleneck that prevents full resource utilization. 

\begin{figure}[ht]
  \centering
  \includegraphics[width=1\linewidth]{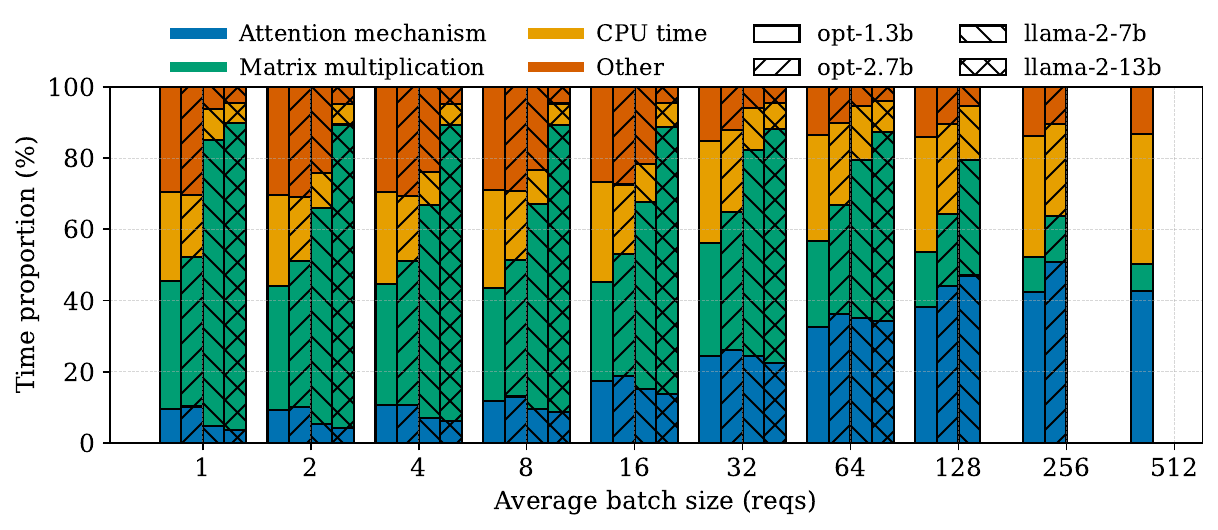}
  \caption{Contribution by kernel to the execution time of the decode steps as the batch size increases across all tested models. Only the kernels regarding matrix multiplications and the attention mechanism are individually labeled, as no other single kernel accounts for more than 20\% of the execution time across any batch size or model. We have also included the amount of time where no GPU kernel is running, labeled as \textit{CPU time}.}
  \label{fig:decode_kernels_distinct_kernels}
\end{figure}

Figure~\ref{fig:decode_kernels_distinct_kernels} shows the contribution of each kernel to the execution time of a single decode step across all tested models and different batch sizes. As expected, matrix multiplications and the attention mechanism dominate execution time. However, as batch size increases, two key trends emerge: the attention mechanism's contribution grows, while the matrix multiplications' impact decreases. For instance, in OPT-1.3B, the attention kernel’s proportion rises from approximately 5\% at small batch sizes to over 40\% at larger ones, whereas matrix multiplications decline sharply from around 50\% to under 10\%. This clearly indicates that the attention mechanism is the primary contributor to execution slowdown in large-batch scenarios. Additionally, CPU computations reach up to 30\% at batch size 512 in OPT-1.3B, highlighting another critical bottleneck that contributes to GPU underutilization.

\begin{figure}[ht]
  \centering
  \includegraphics[width=0.95\linewidth]{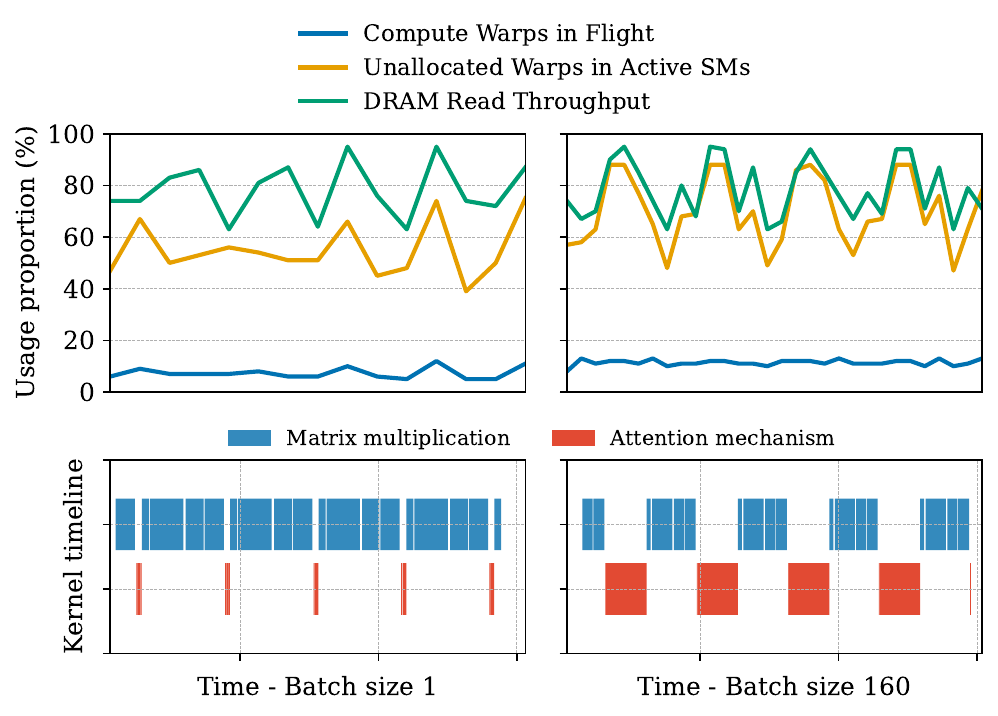}
  \caption{Evolution of the GPU metrics and the kernels regarding the attention mechanism and matrix multiplications during a section of the first decode step of the execution of Llama-2-7B with batch sizes equal to 1 and 160.}
  \label{fig:decode_kernels_timewise}
\end{figure}

We closely examine the kernels associated with the attention mechanism and matrix multiplications. Figure~\ref{fig:decode_kernels_timewise} provides a time-wise representation of their behavior across the execution of multiple layers within a single decode step of the Llama-2-7B model for two batch sizes, along with the corresponding GPU metrics on top. Consistent with previous findings, increasing the batch size results in a greater proportion of execution time spent on the attention mechanism relative to matrix multiplications. More notably, DRAM read saturation occurs exclusively during the execution of the attention kernels, especially at larger batch sizes. This saturation correlates with Warps unallocation. This strongly suggests a DRAM saturation bottleneck inside the attention mechanism that prevents available GPU Warps from being allocated and run in large batch scenarios.

\subsection{Attention Kernel} \label{sec:attention_kernel}

Finally, we analyze the attention kernel in detail, as it becomes the single most critical operation at large batch sizes. Revisiting Figure~\ref{fig:roofline}, we compare the performance and arithmetic intensity of two attention implementations: xFormers~\cite{xFormers2022},  which offers a more memory-efficient implementation using custom CUDA kernels; and FlashAttention \cite{dao2022flashattention}, which further optimizes performance by reducing HBM reads/writes via tiling and recomputation. Our results, extracted from the last decode step of OPT-1.3B at batch size 1 and MAX, show that both algorithms remain firmly within the memory-bound regime across batch sizes, while performance (FLOPS/s) is orders of magnitude lower than the hardware maximum (single precision roofline). The low arithmetic intensity indicates that memory accesses\textemdash that translate into DRAM reads\textemdash consistently exceed the number of compute operations. As illustrated in Figure~\ref{fig:roofline}, the compute-to-memory ratio remains between 0.5 and 1 operations per byte accessed. Moreover, this ratio\textemdash the arithmetic intensity\textemdash remains nearly constant for the two batch sizes, in contrast to matrix multiplication (matmul) kernels, whose arithmetic intensity increases with batch size. This means that attention kernels, unlike matmuls, do not benefit significantly from batching since their performance is fundamentally constrained by DRAM reads. At maximum batch size, attention kernels align with the DRAM bandwidth line, representing the hardware's maximum allowable memory transfer. Given the constant arithmetic intensity, this confirms that the attention performance cannot improve further at larger batch sizes due to DRAM bandwidth saturation. This behavior is not unique to OPT-1.3B,  
as shown in Table~\ref{table:roofline_table}, all tested models exhibit DRAM bandwidth saturation at their maximum batch size.

A deeper analysis of DRAM reads in the attention mechanism reveals inefficient memory access patterns, potentially exacerbated by vLLM’s non-contiguous memory access schema. As shown in Table~\ref{table:hitrate}, the L1 and L2 caches of tested GPUs exhibit consistently low hit rates, averaging no more than 12\% for L1 and 2\% for L2 across all models and batch sizes. These values further decline as the batch size increases, indicating worsening inefficiencies. This poor cache performance significantly reduces overall memory efficiency, decreasing inference performance.

\begin{figure}[ht]
  \centering
  \includegraphics[width=0.95\linewidth]{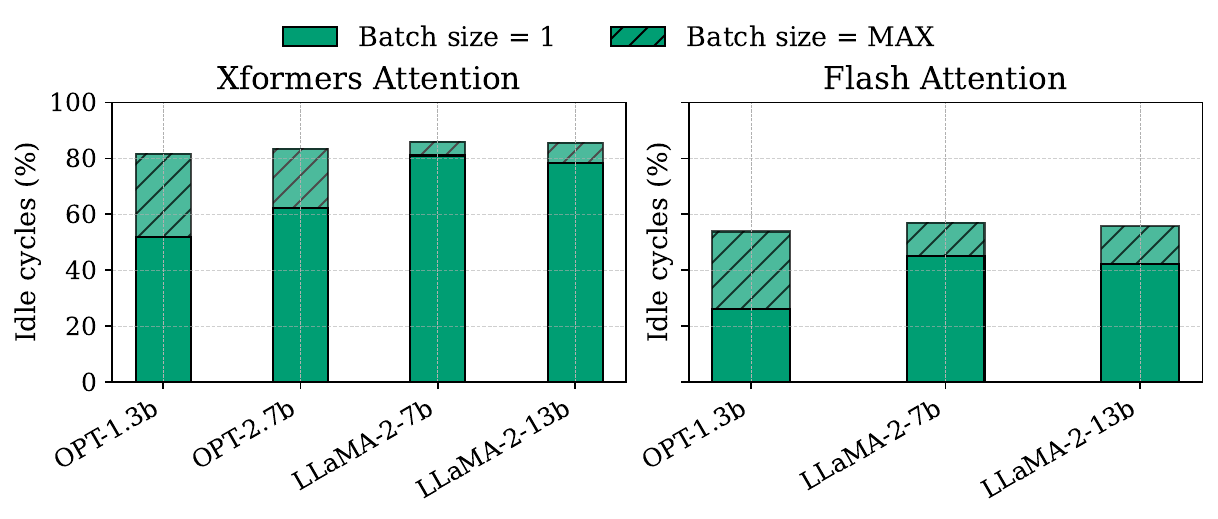}
  \caption{Percentage of the warp cycles issued per instruction that are stalled/idle waiting for data. The results are extracted for all tested models when using no-batch inference and the maximum batch. We use both the xFormers and FlashAttention backends for the attention mechanism, notice that OPT-2.7B model is not compatible with the latter. Every result is the average of the values of the first 5 kernel executions from the last decode step.}
  \label{fig:idlecycles_attention}
\end{figure}

\begin{figure}[ht]
  \centering
  \includegraphics[width=0.95\linewidth]{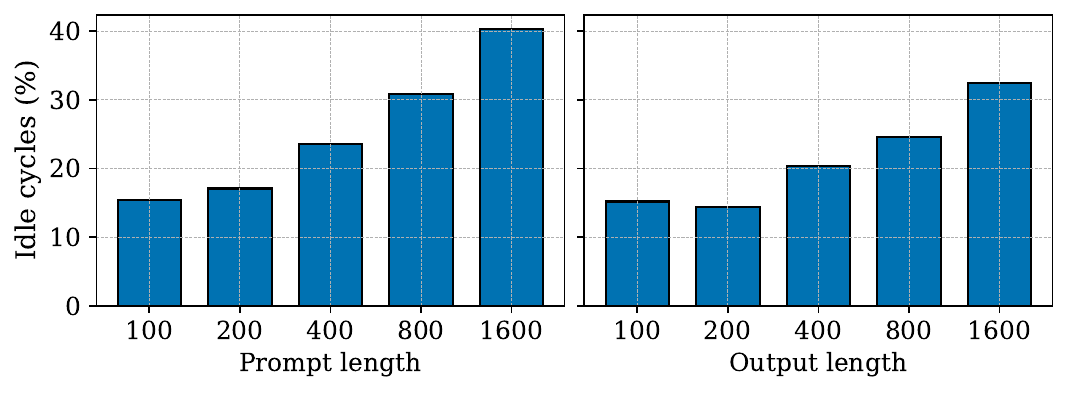}
  \caption{Impact on the percentage of stalled cycles in the FlashAttention kernel in the decode phase when increasing the input and output length separately. The default number of input and output tokens are 100 and 100 respectively. We use the model OPT-1.3B and we average the values from the execution of the attention kernels that run in the first and last decode steps.}
  \label{fig:seqlength_cycles}
\end{figure}

To wrap up this analysis, we examine the proportion of compute cycles spent idly waiting for data. Figure~\ref{fig:idlecycles_attention} presents these values for both attention algorithms across all tested models, comparing batch size 1 with MAX. At maximum batch size, more than 50\% of cycles remain idle due to data-fetching delays, with xFormers being particularly affected, exceeding 80\% idle cycles across all models. Additionally, larger models exhibit higher idle cycles even at batch size 1, highlighting the increasing impact of memory transfers as model size grows. Similarly, sequence length also impacts memory transfer demands, leading to a higher percentage of stalled cycles waiting for data, as shown in Figure~\ref{fig:seqlength_cycles}. As expected, longer prompts have a more pronounced impact than longer output sequences. This is because larger input lengths increase the memory transfer for every decoding step, whereas longer output sequences primarily affect later decoding steps.

Based on these insights and the results from previous sections, we conclude that the throughput plateau is caused by DRAM read saturation in the attention mechanism during the decode phase. As batch size increases, its arithmetic intensity remains constant, and once memory bandwidth limit is reached, performance can no longer improve, making it the primary performance bottleneck.

\begin{table}[htb]
    \renewcommand{\arraystretch}{1.1}
    \centering
    \begin{tabular}{l|l|ccc}
        \multicolumn{2}{@{}c@{}}{} & Batch size & Mem-traffic & Performance \\
        \multicolumn{2}{@{}c@{}}{} & (reqs)   & (byte/s)    & (FLOP/s)    \\ \hline
        \multicolumn{2}{r|}{Rooflines} & - & 1.63E+12 & 2.56E+13 \\ \midrule
        \multirow{8}{3.5em}{Models' achieved values} & \multirow{2}{3.5em}{OPT-1.3B} & 1 & 2.55E+11 & 1.56E+11 \\
        & & 512 & 1.51E+12 & 9.64E+11 \\ \cline{2-5}
        & \multirow{2}{3.5em}{OPT-2.7B} & 1 & 2.17E+11 & 1.31E+11 \\
        & & 256 & 1.56E+12 & 9.42E+11 \\ \cline{2-5}
        & \multirow{2}{3.5em}{Llama-2-7B} & 1 & 1.29E+11 & 7.58E+10 \\
        & & 128 & 1.53E+12 & 9.02E+11 \\ \cline{2-5}
        & \multirow{2}{3.5em}{Llama-2-13B} & 1 & 1.54E+11 & 9.06E+10 \\
        & & 80 & 1.51E+12 & 8.92E+11 \\
    \end{tabular}
    \caption{Roofline results for the xFormers attention algorithm in all tested models when using no-batch inference and their maximum possible batch size. We show the maximum roofline values of the hardware, and the achieved values by the models. Every result is the average of the values of the first 5 kernel executions from the last decode step.}
    \label{table:roofline_table}
\end{table}

\begin{table}[!htb]
    \setlength{\tabcolsep}{3.5pt} 
    \renewcommand{\arraystretch}{1.1}
    \centering
    \begin{tabular}{l|cc|cc|cc|cc}
        & \multicolumn{2}{c|}{OPT-1.3B} & \multicolumn{2}{c|}{OPT-2.7B} & \multicolumn{2}{c|}{Llama-7b} & \multicolumn{2}{c}{Llama-13b} \\ \hline
        Batch size & 1 & 512 & 1 & 256 & 1 & 128 & 1 & 80 \\ \hline
        L1 HR (\%) & 16.49 & 2.62 & 13.84 & 2.43 & 9.40 & 1.55 & 7.70 & 1.61 \\ \hline
        L2 HR (\%) & 1.58 & 1.60 & 1.27 & 1.28 & 0.83 & 0.84 & 0.83 & 0.84 \\
    \end{tabular}
    \caption{Hit rates (HR) of L1 and L2 GPU caches for all tested models when using no-batch inference and MAX batch size. Every result is the average of the values of the first 5 kernel executions from the last decode step.}
    \label{table:hitrate}
\end{table}

\section{Batching Configuration Advisor}\label{sec:advisor}

In this section, we introduce the \textbf{Batching Configuration Advisor (BCA)}, a tool that determines the optimal batch size $B_{opt}$ for LLM serving by jointly considering the throughput plateau and a user-defined latency constraint. BCA is computed offline, prior to online deployment. As previously discussed, increasing the batch size beyond a certain knee-point yields only marginal throughput due to DRAM bandwidth saturation, while increasing GPU memory usage and inter-token latency. BCA addresses this trade-off by selecting a batch size that maximizes throughput, avoids the plateau region, and satisfies a predefined Service Level Objective (SLO). This balanced configuration reduces GPU memory usage and frees up resources for other workloads, improving overall system efficiency.

Formally, Equation~\ref{eq:Bopt} defines $B_{opt}$ as the batch size $B$ that maximizes throughput $T(B)$, subject to two constraints: (i) the latency $L(B)$ must not exceed a specified SLO, and (ii) the throughput relative to the optimal throughput $T(1)*B$ must remain above a user-specified threshold $\epsilon$. Here, $T(B)$ and $L(B)$ represents the throughput and latency observed at batch size $B$, which we determine by benchmarking the model’s performance at each batch size, following the online mode experimental set-up described in Section \ref{sec:methodology}. Both SLO and $\epsilon$ are user-defined parameters.
\begin{equation}\label{eq:Bopt}
\begin{aligned}
B_{\mathrm{opt}}
&= \arg\max_{B}\; T(B) \\
\text{subject to} \quad 
& \left\{
  \begin{array}{l}
    L(B) \le \textsc{SLO}, \\
    \\[-5pt]  
    \frac{T(B)}{B*T(1)}  > \epsilon 
  \end{array}
\right.
\end{aligned}
\end{equation}

\subsection{Evaluation of BCA}

\begin{figure}[ht] 
\centering 
\includegraphics[width=0.95\linewidth]{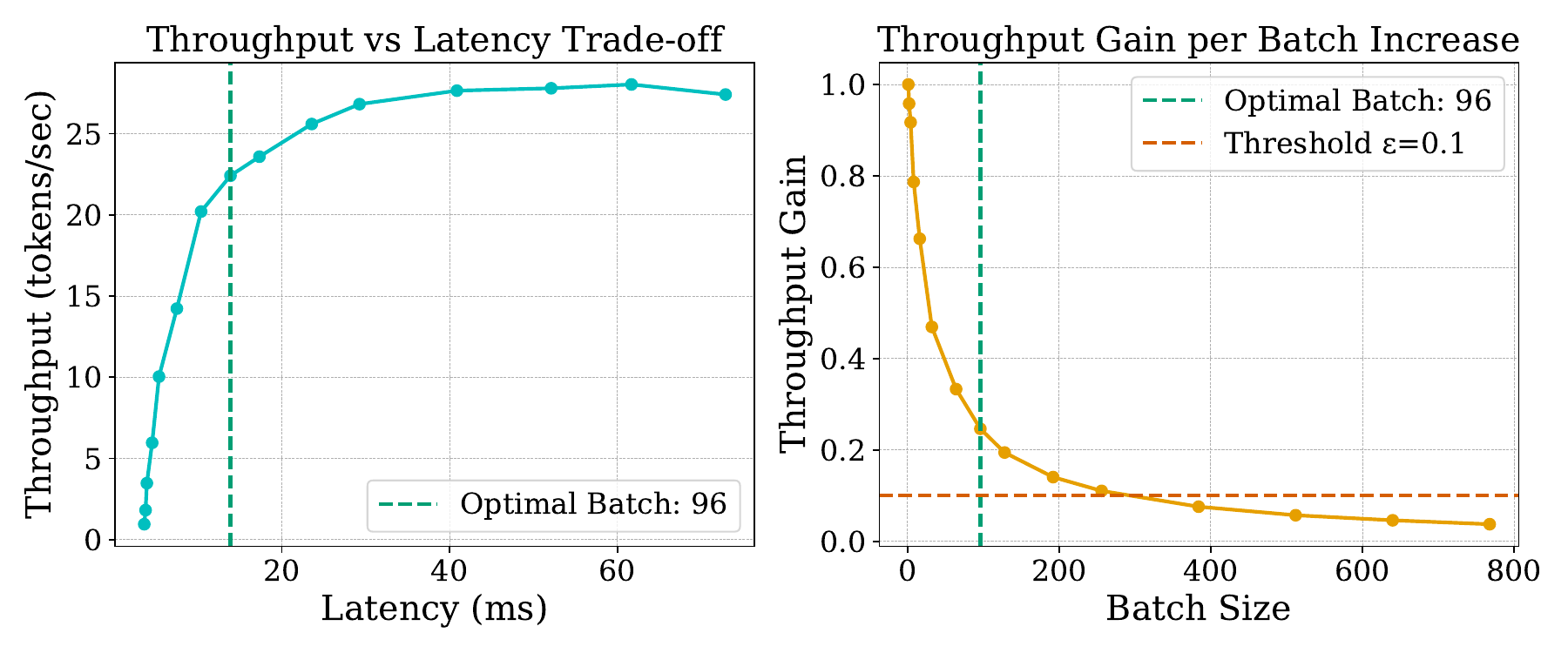} \caption{(Left) Throughput–latency trade-off for OPT-1.3B, highlighting the optimal batch size $B_{opt}$ under strict latency SLO and $\epsilon = 0.1$. (Right) Throughput gain per batch increases relative to ideal linear scaling ($T(1)*B$), highlighting $B_{opt}$ and the threshold $\epsilon = 0.1$.}  
\label{fig:bopt1}
\end{figure}

Table~\ref{table:replication_metrics} presents the results from BCA evaluation across different models with $\epsilon = 0.1$ under two latency SLOs: a strict constraint (2× the latency obtained at batch size 32) and a relaxed constraint (4× the latency obtained at batch size 32). Given these user-defined constraints, BCA identifies the optimal batch size, avoiding diminishing returns in throughput while ensuring that latency constraints are met and GPU memory is efficiently utilized. For instance, in OPT-1.3B, BCA identifies 96 as the optimal batch size under the strict SLO, achieving 83.13\% of the throughput obtained at the maximum batch size while using only 16.32\% of the KV cache. Additionally, this optimal point reduces inter-token latency by a 18.67\%. Figure~\ref{fig:bopt1} further illustrates the balance between throughput and latency for this case. The chosen $B_{opt}$ aligns closely with the knee-point, where throughput stops scaling proportionally with batch size.

\begin{figure}[ht]
  \centering
  \includegraphics[width=0.9\linewidth]{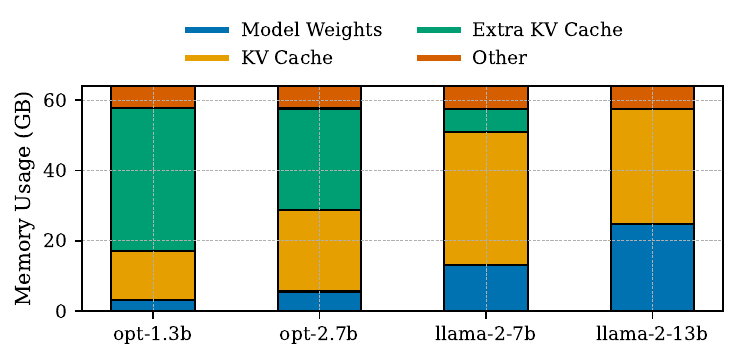}
  \caption{Memory usage distribution for each model size in our 64GB GPU environment, considering $B_{opt}$ under a strict SLO and $\epsilon = 0.1$. By default, vLLM allocates 90\% of available memory, leaving 10\% for the model executor (Other).}
  \label{fig:gpu_memory_distribution_with_bopt_real_data}
\end{figure}

Figure~\ref{fig:gpu_memory_distribution_with_bopt_real_data} visually illustrates the memory savings achieved by BCA under these constraints. The empty KV cache accounts for 63.23\% of the total GPU memory in OPT-1.3B, 45.05\% for OPT-2.7B, and 10.51\% for Llama-2-7B. In contrast, Llama-2-13B requires all available memory to maximize throughput and does not reach the throughput plateau under our hardware evaluation setup. Thus, the effectiveness of BCA heavily depends on model size, available GPU memory, and achievable batch size, with smaller models benefiting the most.

\begin{figure}[ht] 
\centering \includegraphics[width=0.85\linewidth]{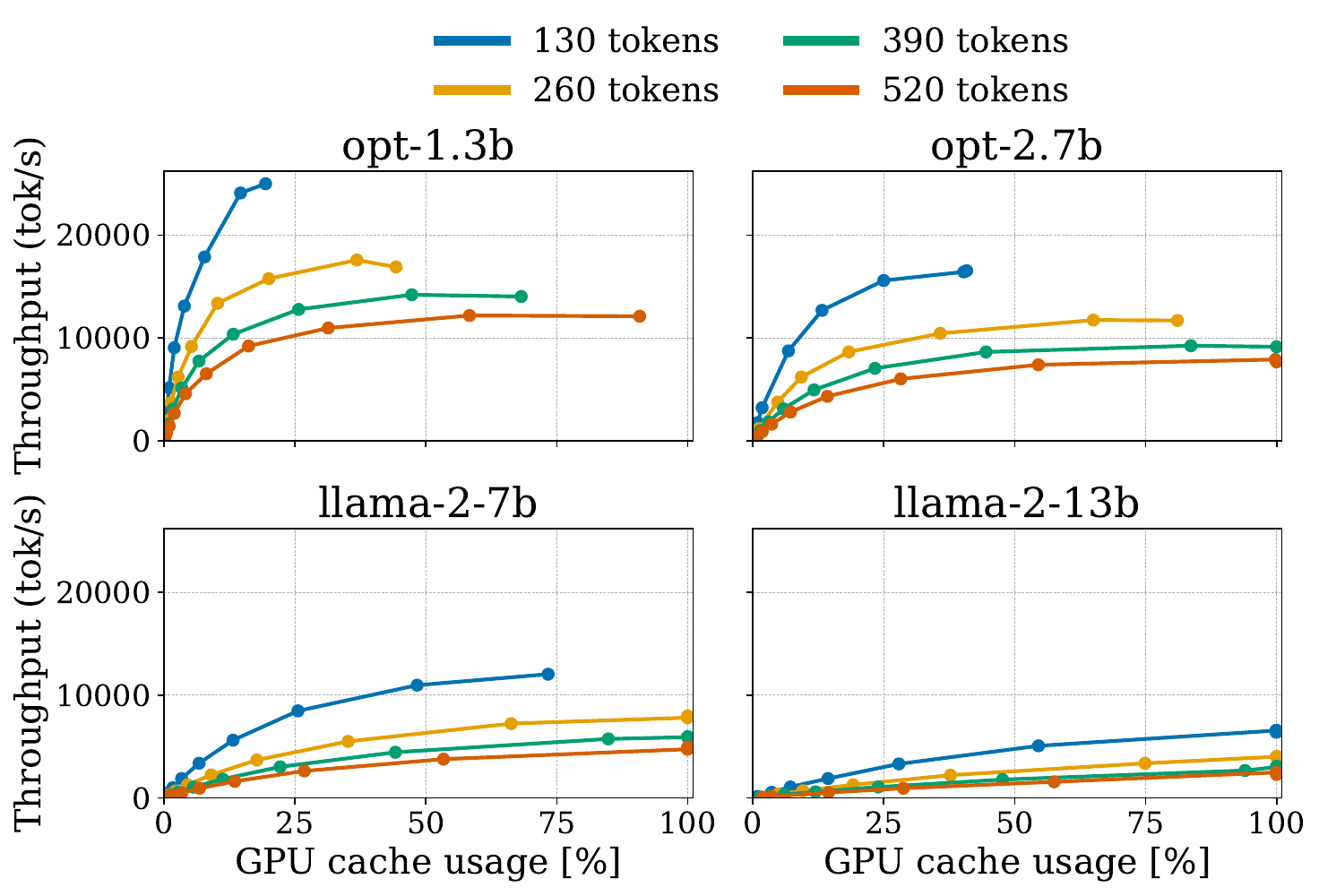} \caption{Throughput versus KV cache usage for batch sizes up to 520 requests, across different output sequence lengths. The average ShareGPT output is 338 tokens, so we evaluate outputs of 130, 260, 390, and 520 tokens.} 
\label{fig:req_length} 
\end{figure}

Finally, Figure~\ref{fig:req_length} illustrates the impact of request sequence length on memory usage. Since GPU DRAM has a fixed capacity, increasing output length causes each batch to consume a larger portion of the KV cache. For example, with OPT-1.3B a batch of 520 requests uses only ~20\% of the KV cache when each request generates 130 output tokens, but consumes over 80\% when each request produces 520 tokens. Thus, while BCA can free significant memory for smaller models with standard output lengths, those gains diminish for unusually long outputs.

\subsection{Model Replication}
In this subsection, we demonstrate how concurrent workloads can utilize the extra GPU memory freed by adopting the BCA-recommended batch size. To this end, we instantiate multiple model servers, each allocated an equal portion of GPU memory, and distribute incoming requests among them. While all replicas experience DRAM read saturation during the attention kernel, they can overlap this period with execution phases of other replicas that are not facing this bottleneck, thereby increasing overall performance. As shown in Table~\ref{table:evaluation_analysis}, DRAM read average throughput remains below 65\% across all models during the decoding phase, indicating room for optimization. Notably, during the CPU time depicted in Figure~\ref{fig:decode_kernels_distinct_kernels}, GPU idle time reaches up to 30\% of the decoding time for OPT-1.3B, presenting further potential for efficiency gain. Additionally, GPU compute resources remain highly underutilized, clearly enabling multiple replicas to run on the same device.

\begin{figure}[ht!]
  \centering
  \includegraphics[width=0.9\linewidth]{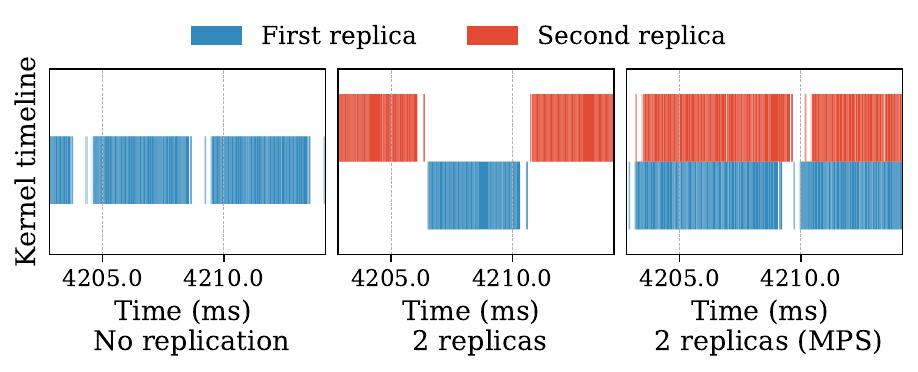}
  \caption{Timeline of a set of decoding steps in OPT-1.3B model under three configurations: no replication, two replicas, and two replicas with NVIDIA MPS. Gaps between decode steps indicate when a replica is idle. In the absence of replication, these gaps represent CPU processing periods during which no GPU kernels are running.}
  \label{fig:replication_timewise}
\end{figure}

To evaluate the impact of replication on LLM serving, we tested two configurations. The first one executes decode steps from replicas in a first-come, first-served manner (FCFS), while the second runs them in parallel using NVIDIA Multi-Process Service (MPS). Figure~\ref{fig:replication_timewise} shows the behavior of these two configurations. In the first approach, replicas improve GPU resource utilization by filling GPU gaps during CPU computations, whereas with MPS, parallel execution of kernel operations enables better resource utilization throughout all the entire decoding process. Given these advantages, we adopt MPS as our replication strategy, and all subsequent results follow this configuration.

Table~\ref{table:replication_metrics} presents the final results of replication over the BCA-recommended batch size. Following the previous subsection, we determine $B_{opt}$ using $\epsilon = 0.1$ and evaluate both strict and relaxed latency constraints. Given the memory required for $B_{opt}$, we incrementally increase the number of replicas until the GPU memory is fully utilized. For Llama-2-7B and Llama-2-13B, the identified $B_{opt}$ values do not allow replication, as replicating the memory required for $B_{opt}$ exceeds available GPU memory. Overall, the results confirm that replication effectively utilizes the GPU memory freed by BCA. Under both strict and relaxed SLOs, throughput improves compared to a single replica, even surpassing the one from maximum batch size (MAX) while using less KV cache. For OPT-1.3B, replication under the relaxed configuration achieves a 34\% throughput increase over MAX, whereas for OPT-2.7B, the increase reaches 13\%. Regarding latency, replication increases inter-token latency by an average of 28\% across both models compared to $B_{opt}$. While this ITL increase remains significantly lower than of MAX, it is still a factor to be considered. However, end-to-end latency decreases, indicating that parallelizing decoding steps across replicas slows down each individual step while increasing overall output token generation. To contextualize these results, Table~\ref{table:replication_metrics} also includes the performance of chunked prefill with MAX batch size, a serving optimization technique introduced in Section~\ref{sec:relatedwork}. As shown, replication achieves comparable or even superior performance compared to chunked prefill in both models. Future work should explore combining replication with chunked prefill to further optimize serving performance and assess additional potential benefits in multi-replica setups.

A closer examination of the GPU metrics in Table~\ref{table:replication_metrics} confirms that replication enhances resource utilization. GPU compute activity increases, as indicated in the \textit{Compute Warps in Flight} metric. More notably, there is a significant rise in average DRAM read, demonstrating that replication helps mitigate DRAM hardware saturation in the attention mechanism. Interestingly, in contrast, the MAX batch size achieves similar DRAM read values to those observed at $B_{opt}=96$ in both models, without any increase. This improvement is primarily due to overlapping GPU gaps during CPU computations. As shown, the CPU time is reduced by an average of 78\% across both models when using two replicas. This also explains the limited throughput increase when scaling from 2 to 4 replicas in OPT-1.3B\textemdash since CPU time has been already significantly reduced with two replicas, further replication provides diminishing returns in performance gains.

\begin{table*}[htbp]
    \renewcommand{\arraystretch}{1.2}
    \centering
    \setlength{\tabcolsep}{5pt}
    \begin{tabular}{l|lc|>{\centering\arraybackslash}m{1.1cm}>{\centering\arraybackslash}m{1.1cm}>{\centering\arraybackslash}m{1.1cm}>{\centering\arraybackslash}m{1.1cm}|>{\centering\arraybackslash}m{1.1cm}>{\centering\arraybackslash}m{1.1cm}>{\centering\arraybackslash}m{1.1cm}}
        & & & \multicolumn{4}{c|}{\textbf{Serving Metrics}} & \multicolumn{3}{c}{\textbf{GPU Metrics}} \\ \cline{4-10}
        \textbf{Model} & \textbf{Batch Size} & \textbf{Replicas} & \textbf{Throughput} & \textbf{ITL} & \textbf{E2E} & \textbf{KV Cache Usage} & \textbf{Compute Warps in Flight} & \textbf{DRAM Read} & \textbf{CPU Time} \\
        & (reqs) & (\#) & (tokens/ms) & (ms) & (s) & (\%) & (\%) & (\%) & (\%) \\ \midrule
        
        \multirow{7}{*}{\textbf{OPT-1.3B}} & MAX & 1 & 10.97 & 73.77 & 30.39 & 97.22 & 8.17 & 46.66 & 36.51 \\
        & MAX (with chunked prefill) & 1 & 11.86 & 65.30 & 26.75 & 96.71 & 9.02 & 49.23 & 35.34 \\ \cline{2-10}
        & \multirow{3}{*}{$B_{opt}=96$ - Strict SLO, $\epsilon = 0.1$} & 1 & 9.12 & 13.78 & 43.42 & 15.87 & 6.84 & 47.14 & 23.11 \\
        & & 2 & 12.31 & 18.98 & 30.64 & 30.46 & 10.91 & 66.51 & 5.64 \\
        & & 4 & 13.17 & 31.52 & 24.80 & 71.07 & 13.80 & 77.34 & 1.03 \\ \cline{2-10}
        & \multirow{2}{*}{$B_{opt}=256$ - Relaxed SLO, $\epsilon = 0.1$} & 1 & 10.87 & 29.26 & 34.85 & 37.9 & 7.76 & 47.76 & 29.10 \\
        & & 2 & 14.67 & 37.07 & 22.11 & 74.47 & 11.67 & 67.73 & 7.58 \\ \midrule
        
        \multirow{6}{*}{\textbf{OPT-2.7B}} & MAX & 1 & 7.43 & 61.60 & 46.57 & 96.44 & 26.80 & 58.92 & 22.74 \\
        & MAX (with chunked prefill) & 1 & 8.32 & 53.40 & 40.43 & 95.81 & 28.78 & 60.94 & 22.10 \\ \cline{2-10}
        & \multirow{2}{*}{$B_{opt}=96$ - Strict SLO, $\epsilon = 0.1$} & 1 & 6.17 & 20.35 & 62.63 & 28.38 & 20.4 & 58.70 & 15.74 \\
        & & 2 & 7.73 & 30.64 & 46.44 & 56.42 & 27.09 & 77.50 & 2.62 \\ \cline{2-10}
        & \multirow{2}{*}{$B_{opt}=128$ - Relaxed SLO, $\epsilon = 0.1$} & 1 & 6.60 & 25.06 & 57.43 & 33.41 & 21.91 & 58.67 & 16.08 \\
        & & 2 & 8.38 & 36.55 & 41.39 & 77.15 & 29.21 & 77.36 & 3.11 \\
        
    \end{tabular}
    \caption{Serving and GPU metrics for OPT-1.3B and OPT-2.7B, comparing the maximum allowed batch size (MAX) and the recommended batch size from BCA ($B_{opt}$). For $B_{opt}$, replication is employed to maximize resource utilization—up to four replicas for OPT-1.3B and two for OPT-2.7B. Throughput measures the rate of token processing, while CPU time refers to periods where no GPU kernels are active. GPU metrics are extracted through the NVIDIA Nsight Systems (2023.2.3) tool.}
    \label{table:replication_metrics}
\end{table*}

\section{Discussion}
\label{sec:discussion}

In this work, we identify the GPU performance bottlenecks responsible for throughput plateaus in large-batch LLM inference. We find that the arithmetic intensity of attention kernels remains nearly constant as batch size increases, leading to DRAM bandwidth saturation at larger batches. This DRAM saturation is the principal factor behind the performance slowdown beyond a batch-size knee point, leaving most GPU compute resources underutilized. Additionally, CPU overhead grows with batch size\textemdash reaching up to 30\% of the total execution time in some cases,  further limiting scalability. Our study specifically focuses on the inference behavior of smaller LLMs that fit within a single GPU, allowing us to explore large batch sizes without multi-GPU communication overhead. While these findings are highly relevant for optimizing LLM serving for smaller models, we leave as future work the exploration of bottlenecks in larger models, where inter-GPU communication overheads and increased memory constraints will likely play significant roles.

The proposed Batching Configuration Advisor (BCA) determines a batch size for LLM serving that maximizes throughput while adhering to user-defined latency constraints. Unlike existing approaches that allocate full GPU memory by default, BCA allocates only the memory necessary to achieve the optimal batch size. The advisor estimates the optimal configuration in an offline manner, assuming that all requests arrive simultaneously. However, real-world serving scenarios exhibit variable request patterns. In such scenarios, BCA serves as an upper-bound estimator, as it reflects the best-case performance achievable under ideal batch conditions. Future work should extend BCA to adapt the batch size in an online manner, dynamically adjusting memory allocations based on incoming request patterns.

Finally, we demonstrate that the resources freed by BCA can be leveraged for concurrent workloads. Specifically, we evaluate the impact of running multiple instances in parallel on the same GPU. Our replication strategy increases overall throughput by 33.72\% for OPT-1.3B (with 4 replicas) and by 12.78\% for OPT-2.7B (with 2 replicas), primarily by mitigating GPU idle cycles caused by CPU bottlenecks at large batch sizes. This is especially important for multi-model serving in shared cloud environments. For future work, we suggest extending replication strategies to heterogeneous workloads with opposite resource usage patterns to LLM inference (e.g., high compute utilization but low memory demands). Another promising direction is replicating larger models across multiple GPUs, requiring solutions for inter-GPU communication bottlenecks and distributed memory constraints. Together, these strategies pave the way for a more efficient and holistic use of GPU resources for LLM serving, transforming idling capacity into opportunities for concurrent processing and faster inference.

\section{Conclusion}

In this work, we conducted an in-depth GPU analysis to identify performance bottlenecks that cause the throughput plateaus in large-batch inference. Our findings challenge the prevailing assumption that large-batch inference transitions into a compute-bound regime and fully utilizes compute resources; instead, we demonstrate that DRAM bandwidth saturation remains the primary bottleneck, leaving significant compute resources underutilized. To address this inefficiency, we propose a Batching Configuration Advisor (BCA), which determines the optimal batch size and prevents unnecessary GPU memory allocation. Additionally, we show that freed memory from BCA can be leveraged for concurrent workloads via GPU sharing techniques (time-sharing and MPS). Specifically, we evaluate replicating smaller LLMs and running multiple instances to improve GPU utilization by overlapping operations, mitigating DRAM saturation, and improving serving throughput. Our findings challenge conventional LLM inference assumptions and provide practical strategies for optimizing GPU efficiency through optimal batching and GPU sharing.

\newpage 
\section*{Acknowledgments}

This work has been partially financed by grant agreement EU-HORIZON GA.101095717 and by the EU-HORIZON MSCA programme under grant agreement EU-HORIZON MSCA GA.101086248. Also, it has been partially financed by Generalitat de Catalunya (AGAUR) under grant agreement 2021-SGR-00478, by Severo Ochoa Center of Excellence CEX-2021-001148-S-20-3, and by the Spanish Ministry of Science (MICINN), the Research State Agency (AEI) and European Regional Development Funds (ERDF/FEDER) under grant agreement PID2021-126248OB-I00, MCIN/AEI/10.13039/ 501100011033/ FEDER, UE.

\bibliographystyle{ACM-Reference-Format}
\bibliography{ms}


\end{document}